\documentclass[aps,prl,twocolumn,superscriptaddress]{revtex4}

\usepackage{graphicx}
\usepackage{amsmath}
\usepackage{amssymb}
\usepackage{amsthm}

\begin{document}

\title{Growing scale-free simplices}

\author{K. Kovalenko$^{*}$}
\affiliation{Moscow Institute of Physics and Technology (National Research University), 9 Institutskiy per., Dolgoprudny, Moscow Region, 141701, Russian Federation}

\author{I. Sendi\~na-Nadal$^{*}$}
\affiliation{Complex Systems Group \& GISC,  Universidad Rey Juan Carlos, 28933 M\'ostoles, Madrid, Spain}
\affiliation{Center for Biomedical Technology, Universidad Polit\'ecnica
de Madrid, 28223 Pozuelo de Alarc\'on, Madrid, Spain}

\author{N. Khalil$^{*}$}
\affiliation{ESCET \& GISC, Universidad Rey Juan Carlos, 28933 M\'ostoles, Madrid, Spain}

\author{A. Dainiak}
\affiliation{Moscow Institute of Physics and Technology (National Research University), 9 Institutskiy per., Dolgoprudny, Moscow Region, 141701, Russia}

\author{D. Musatov}
\affiliation{Moscow Institute of Physics and Technology (National Research University), 9 Institutskiy per., Dolgoprudny, Moscow Region, 141701, Russia}
\affiliation{Russian Academy of National Economy and Public Administration, pr. Vernadskogo, 84, 119606 Moscow, Russia}
\affiliation{Caucasus Mathematical Center at Adyghe State University, ul. Pervomaiskaya, 208, Maykop, Russia}

\author{A.M. Raigorodskii}
\affiliation{Moscow Institute of Physics and Technology (National Research University), 9 Institutskiy per., Dolgoprudny, Moscow Region, 141701, Russian Federation}
\affiliation{Caucasus Mathematical Center at Adyghe State University, ul. Pervomaiskaya, 208, Maykop, Russia}
\affiliation{Mechanics and Mathematics Faculty, Moscow State University, Moscow, Russia}
\affiliation{Institute of Mathematics and Computer Science, Buryat State University, Ulan-Ude, Russia}

\author{K. Alfaro-Bittner}
\affiliation{Unmanned Systems Research Institute, Northwestern Polytechnical University, Xi'an 710072, China}
\affiliation{Departamento de F\'isica, Universidad T\'ecnica Federico Santa Mar\'ia, Av. Espa\~na 1680, Casilla 110V, Valpara\'iso, Chile}

\author{B. Barzel}
\affiliation{Department of Mathematics, Bar-Ilan University, Ramat-Gan, 5290002, Israel}
\affiliation{Gonda Multidisciplinary Brain Research Center, Bar-Ilan University, Ramat-Gan, 5290002, Israel}

\author{S. Boccaletti}
\affiliation{Moscow Institute of Physics and Technology (National Research University), 9 Institutskiy per., Dolgoprudny, Moscow Region, 141701, Russia}
\affiliation{Unmanned Systems Research Institute, Northwestern Polytechnical University, Xi'an 710072, China}
\affiliation{CNR - Institute of Complex Systems, Via Madonna del Piano 10, I-50019 Sesto Fiorentino, Italy}

\begin{abstract}
 The past two decades have seen significant successes in our understanding of complex networked systems, from the mapping of real-world social, biological and technological networks to the establishment of generative models recovering their observed macroscopic patterns. These advances, however, are restricted to pairwise interactions, captured by dyadic links, and provide limited insight into higher-order structure, in which a group of several components represents the basic interaction unit. Such multi-component interactions can only be grasped through simplicial complexes, which have recently found applications in social and biological contexts, as well as in engineering and brain science. What, then, are the generative models recovering the patterns observed in real-world simplicial complexes? Here we introduce, study, and characterize a model to grow simplicial complexes of order two, i.e. nodes, links and triangles, that yields a highly flexible range of empirically relevant simplicial network ensembles. Specifically, through a combination of preferential and/or non preferential attachment mechanisms, the model constructs networks with a scale-free degree distribution and an either bounded or scale-free generalized degree distribution – the latter accounting for the number of triads surrounding each link. Allowing to analytically control the scaling exponents we arrive at a highly general scheme by which to construct ensembles of synthetic complexes displaying desired statistical properties.

\end{abstract}
\maketitle

{\it * These Authors equally contributed to the Manuscript}

\vskip 0.5 truecm

All the beauty, richness and harmony in the emergent dynamics of a complex system largely depend on the specific way in which its elementary components interact.
The last twenty years have seen the birth and development of the multidisciplinary field of {\it Network Science}, wherein a variety of systems in physics, biology, social sciences and engineering have been modelled as networks of coupled units, in the attempt to unveil the mechanisms underneath their observed functionality \cite{doro2002,newman2003,boccaletti06,Estrada2011,Boccaletti2014,Latora2017}.

But the fundamental limit of  such a representation is that networks capture only pairwise interactions, whereas the function of many real-world systems not only involves dyadic connections, but rather is the outcome of collective actions at the level of groups of nodes. For instance, in ecological systems, three or more species compete for food or territory  \cite{levine}. Similar multi-component interactions appear in functional \cite{petri2,sizemore2016,Lee2012,Petri2014,Lord2016} and structural \cite{sizemore} brain networks, protein interaction networks \cite{estrada2}, semantic networks \cite{sizemore2}, multi-Authors scientific collaborations \cite{patania}, offline and online social networks \cite{freeman,Andjelkovic},  trigenic interactions in gene regulatory networks \cite{Uzi,kuzmin}, and spreading of contagious diseases due to multiple, simultaneous, contacts \cite{centola}.

Simplicial complexes (SCs), being structures formed by {\it simplices} of different dimensions (nodes, links, triangles, tetrahedra, etc..), can effectively map the relationships between any number of components.
Originally introduced over two decades ago \cite{aleksandrov}, SCs are becoming increasingly relevant thanks to the enhanced resolution of current data sets and the recent advances in data analysis techniques  \cite{salnikov,Sizemore2019}. As real data is being accumulated, we encounter a theoretical challenge: how to synthesize SCs that faithfully reproduce the observed structural features. While significant progresses were already made in extending to SCs static graph models (such as random graphs \cite{costa,Iacopini2019}, or the configuration model\cite{courtney}, or activity driven models \cite{petri3}), the same attention has not yet been paid to the study of growth models for SCs, despite the fact that in many circumstances the network is the result of a growing process, such as in the case of scientists collaborating with and citing each other. A step forward has been made in the field of network geometry \cite{boguna2020}
and its relation with quantum networks \cite{Bianconi2016,Courtney2017} or complex materials \cite{tadic2018}, where random graph models have been developed by aggregating SCs as fundamental building blocks.

\begin{figure}
\centering
\includegraphics[width=0.5\textwidth]{{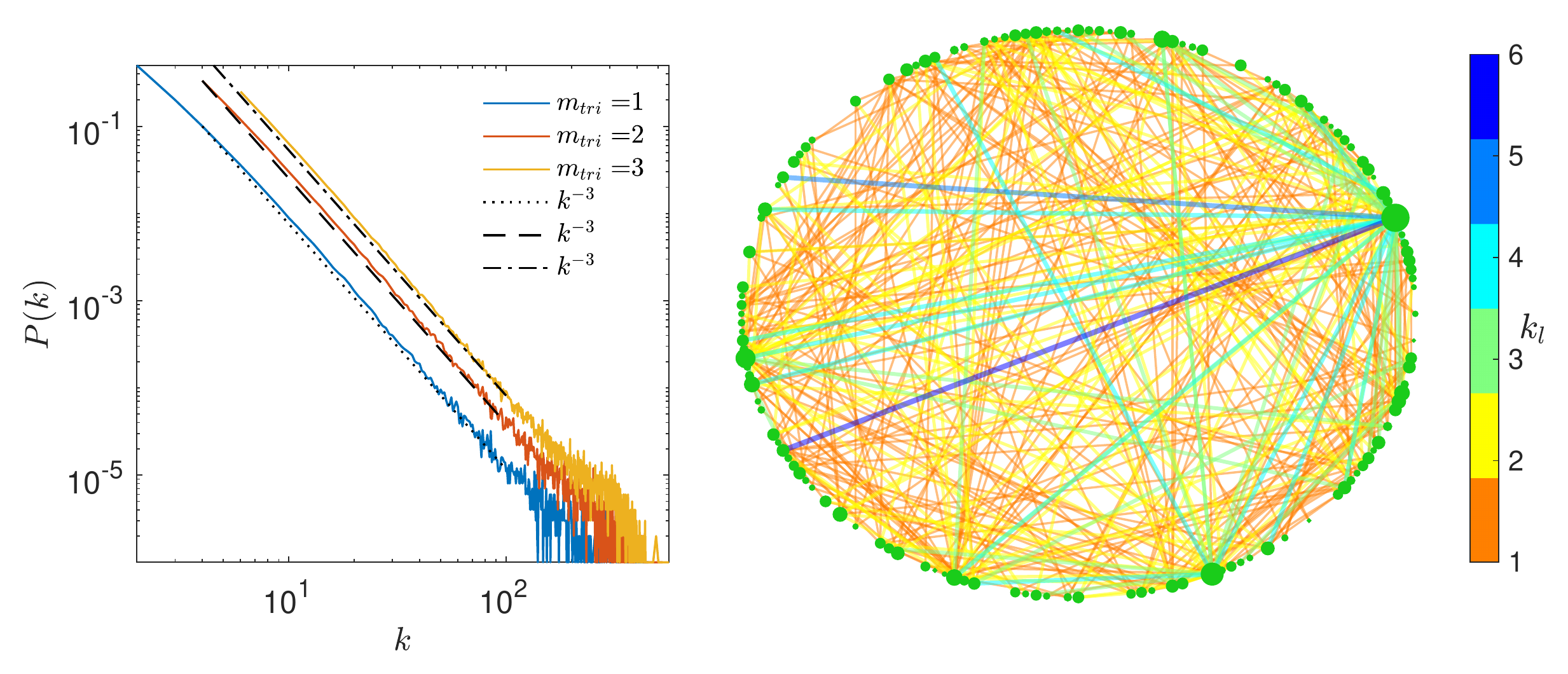}}
\includegraphics[width=0.5\textwidth]{{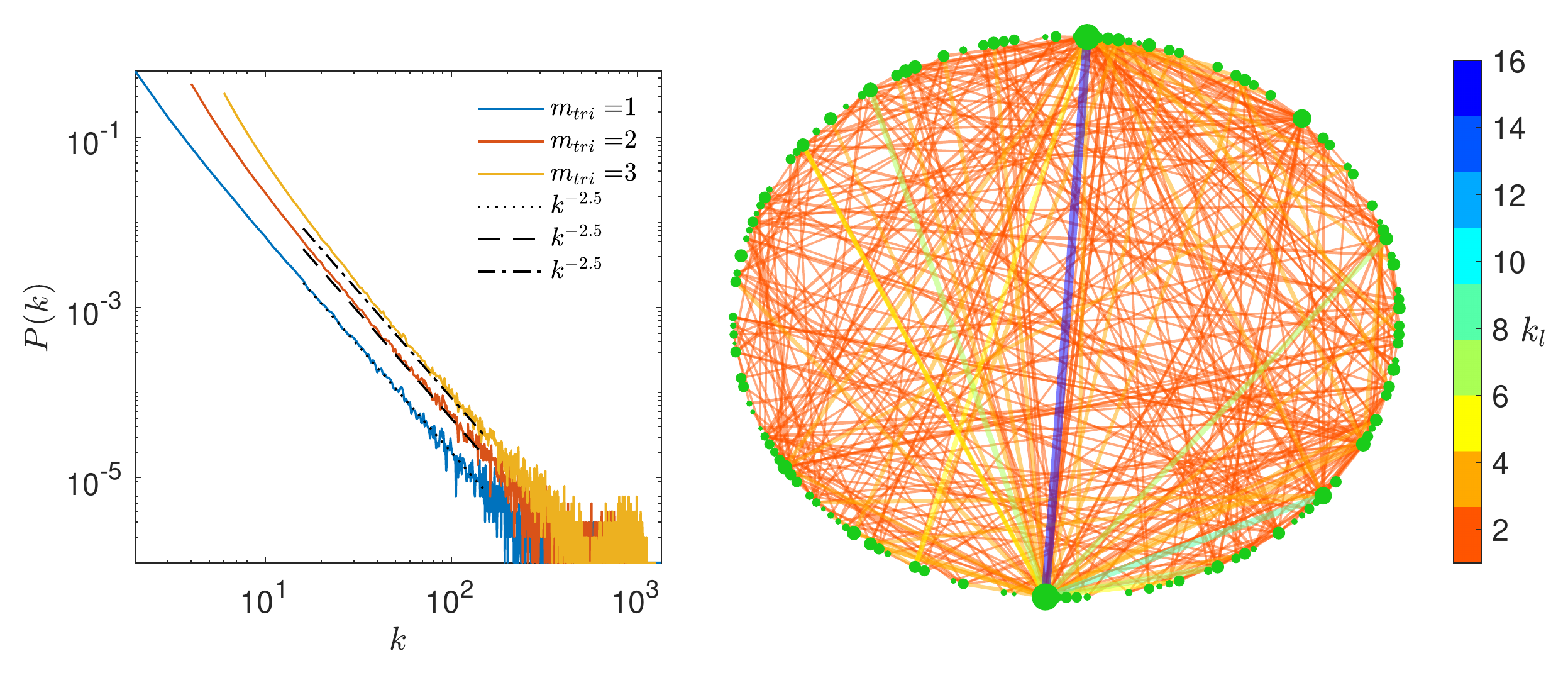}}
\caption{{\bf The structures generated in the preferential and the non preferential case.} The model without preferential attachment (upper panels) and that with preferential attachment (lower panels).
Left panels: log-log plot of the degree distribution $P(k)$ of the resulting networks for different values of $m_{tri}$ (see color code in the legend). Data  are obtained as an ensemble average over 100 different realizations of a network with size $N = 10,000$ nodes.
Dashed lines in the upper (lower) panel correspond to the analytical predictions given by Eqs.~(\ref{eq:ranPk}) [(\ref{eq:powlaw})].
Right panels: schematic visualization of generated networks with $N=200$ and $m_{tri}=1$.  The size of the nodes is proportional to 50 times the square root of the corresponding entry in the eigenvector centrality \cite{Bonacich}, the width of each link is  proportional to $\sqrt{k_{ij}}$, and the color of the links encodes $k_l$ as reported in the bars at the right of both panels.}
\label{fig1}
\end{figure}

In this Letter we introduce and fully characterize several methods able to grow SCs of order two, i.e. structures made of nodes, links and triangles, making use of  preferential and non preferential rules, as well as combinations of both.
The resulting SCs are characterized by two distributions, the classic degree distribution $P(k)$, capturing the fraction of nodes with degree $k$, and the generalized degree distribution $P(k_l)$, where $k_l$ characterizes the number of triangles supported by each link $l = (i,j)$. We show that our generative model always yields a power-law scaling in $P(k)$, recovering the ubiquitously observed scale-free property \cite{barabara,barabara2}, while, at the same time, allowing full control over $P(k_l)$, i.e. bounded or scale-free with any desired scaling exponent. Indeed, $P(k_l)$ has been shown to play a crucial role in the emergence of collective behavior, such as synchronization \cite{boccasynch}.

The purpose is to grow a network of $N$ nodes featuring a transitivity coefficient $T=1$ \cite{barrat2000,newman}, thus implying that each link is part of
a connected triplet of nodes (a triad) which forms a triangle. This is done by starting at $t=0$ with an elementary network seed, an initial clique (an all-to-all connected network) of size $N_0\ll N$. Then, at each successive times $t=1,2,3,..., N-N_0$ a new node is added to the graph. The added node selects $m_{tri}$ already existing links, and forms connections with the $2m_{tri}$ nodes located at the ends of such links, thus generating $m_{tri}$ new triangles (when $m_{tri}>1$ a condition is enforced that the $m_{tri}$ selected links are not pairwise adjacent, to avoid multiple links from the added node to single existing nodes in the graph).
Such a procedure can be conducted with or without adopting a preferential attachment rule. In the first case, the probability that the node added at time $t$ selects the specific link $(i,j)$ to form its connections is taken to be $P_{ij} (t) = \frac{1}{N_{0}\left(N_{0}-1\right) / 2 +2m_{tri} (t-1)}$, i.e. the $m_{tri}$ links are randomly selected among all those which already exist in the graph at time $t-1$ with equal probability (and therefore without applying any preferential rule).
In the second case, instead, one has that $P_{ij} (t) = \frac{k_{ij}(t-1)}{\sum_{i,j} k_{ij}(t-1)}$, which implies that the larger is the number of triangles a given edge $ij$ is part of at time $t-1$ the larger the probability for that edge of being selected to form a new triangle with the added node.

The left top and bottom panels of Figure 1 report the degree distribution $P(k)$ of the networks (of size $N=10,000$ nodes) generated by the two methods. One immediately sees that in both cases the graphs feature a clear power-law scaling, i.e. the scale-freeness which is indeed characterizing the vast majority of real world networks \cite{doro2002,newman2003,boccaletti06,Estrada2011,Boccaletti2014,Latora2017,barabara,barabara2}. In the two right panels of the same Figure we report, instead, a visualization of a typical synthesized network with $N=200$ and $m_{tri}=1$.
We also measure the generalized degree $k_l$. Given a graph of $N$ nodes and its adjacency matrix $A$ (the $N {\text x} N$ matrix with entries $a_{ij} =1$ if nodes $i$ and $j$ are connected by a link, and $a_{ij} =0$ otherwise), the generalized degree $k_{ij}$ of the link between node $i$ and $j$ is the $(i,j)$-entry of the matrix $A \oslash (A^2)$ (with the symbol $\oslash$ standing for the Hadamard product).
While both generated networks are essentially heterogeneous in the node degree, the properties of $k_l$ appear instead to be very different to one another. In particular, the non preferential case leads to a much more restricted range of $k_l$ values as compared to that generated by the preferential rule (see the two color bars at the right of the panels). Moreover, the structure obtained with no preferential attachment is homogeneous in terms of $k_l$ (almost all the six colors, each one representing a given value of $k_l$, are visible). On the opposite, the preferential rule generates a SC with a high heterogeneity in $k_l$: almost all links are red (they have the lowest as possible value of $k_l$), and only one link (the blue one) displays a value of $k_l$ equal to the maximum in the distribution.
These features are  more quantitatively visible in Figure 2, which reports the distribution $P(k_l)$ of the generalized degree $k_l$ for the two cases. Comparing panels (a) and (b) of Fig. 2, one immediately realizes that while $P(k_l)$ is exponentially decaying when the growth is realized with no preferential attachment (Fig. 2a), its scaling is a clear power-law in the presence of preferential attachment. A relevant conclusion at this stage is that the two cases are imprinting a completely different topology in the triangular structures (reflected by completely different scaling properties in the distribution of the generalized degree $k_l$).

We now furnish a full analytic treatment, and provide the rigorous expressions for the distributions $P(k)$ and $P(k_l)$.  We start with the case of no preferential attachment, and we call $N(k,t)$ the number of nodes with degree $k$ at time $t$. Its rate equation reads
$\frac{dN(k,t)}{dt}=\frac{2m_{tri}}{\sum_k kN(k,t)}\left[-kN(k,t)+(k-1)N(k-1,t)\right] +\delta_{k,2m_{tri}}$,
where $\frac{dN(k,t)}{dt}\equiv N(k,t+1)-N(k,t)$, and $\delta$ is the Kronecker delta-function.
For $N(t)\equiv \sum_{k}N(k,t)\simeq t$, one seeks a solution of the form $N(k,t)=tP(k)$, where $P(k)$ is assumed to be time independent. Since the total number of edges is approximately $2m_{tri}t$, one has $\sum_kkN(k,t)\simeq 4m_{tri}t$, and the rate equation for $P(k)$ becomes
$P(k)=\frac{k-1}{k+2}P(k-1)+\frac{2}{k+2}\delta_{k,2m_{tri}}$. Such a latter equation is the same as Eq. (6) of Ref. \cite{Boccaletti2007}, and its solution (for $ k\ge 2m_{tri}$) is
\begin{eqnarray}
  P(k)=\frac{4m_{tri}(2m_{tri}+1)}{k(k+1)(k+2)} \sim k^{-3},
  \label{eq:ranPk}
\end{eqnarray}
which perfectly fits the data of Figure 1 (see the dotted, dashed-dotted, and dashed black lines in the top-left panel).

\begin{figure}
\centering
\includegraphics[width=0.5\textwidth]{{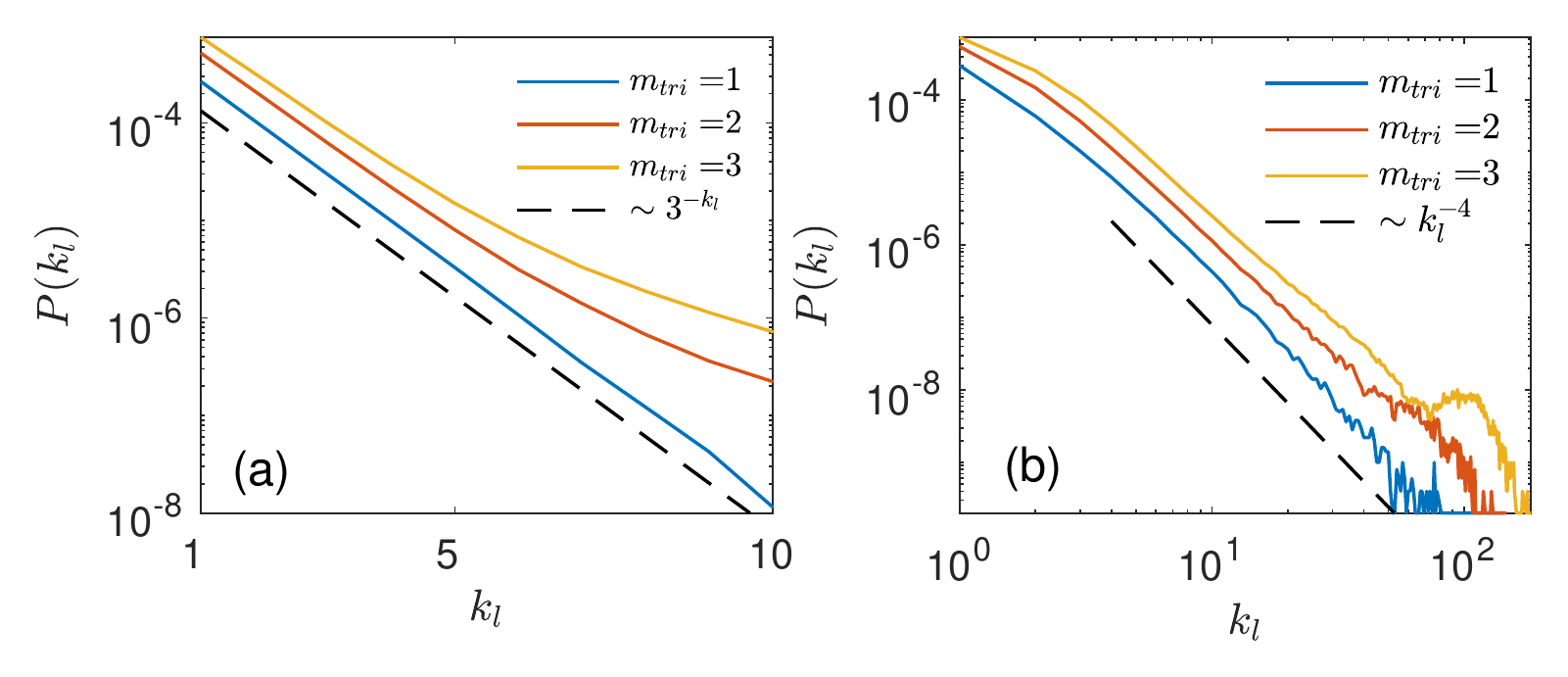}}
\caption{{\bf The distribution of the generalized degree $k_l$.}  The distribution $P(k_l)$ vs. the generalized degree $k_l$ obtained by growing
a network of size $N = 10,000$ with the non preferential model [panel (a)] and with the preferential model [panel (b)]. The data refer to an ensemble average
over 100 different realizations of the growth process. Notice that panel (a) is in log-linear scale, whereas panel (b) is in log-log scale.
Legends in both panels report the color code for the number $m_{tri}$ used in the growth process. The dashed line in panel (a) is used for exponential solution given by Eq.~(\ref{eq:ranPDeltaeNP}), while the dashed line in panel (b) is used for the power-law solution given by Eq.~(\ref{expolaw}). \label{fig2}}
 \end{figure}

Then one can consider $N_e(k_l,t)$ as the number of edges participating in $k_l$ triangles at time $t$. Its rate equation is
$\frac{dN_e(k_l,t)}{dt}=-m_{tri}\frac{N_e(k_l,t)}{N_e(t)}+m_{tri}\frac{N_e(k_l-1,t)}{N_e(t)} +2m_{tri}\delta_{k_l,1}\pm \dots$,
where $N_e(t)\simeq 2m_{tri}t$. For the distribution of the generalized degree $k_l$, one has that $P(k_l)=N_e(k_l,t)/N_e(t)$, and the recursive equation is $P(k_l)=\frac{1}{3}P(k_l-1)+\frac{2}{3}\delta_{k_l,1}$, admitting the following solution:
\begin{equation}
  P(k_l)=\frac{2}{3^{k_l}},\qquad k_l\ge 1.
  \label{eq:ranPDeltaeNP}
\end{equation}
 The exponential function (\ref{eq:ranPDeltaeNP}) is reported as a dashed line in panel (a) of Fig. \ref{fig2},  and one can see that the fit with numerical simulations is rather good, especially for the case $m_{tri}=1$.

The analytic treatment of the preferential attachment case is far more complicated, as it implies the demonstration of a couple of Theorems and the extensive use of a known Lemma. We limit here ourselves to furnish the main results without reporting all the (sometimes cumbersome) formal mathematical steps, whereas the interested reader can find the full details within a large section of our Supplemental Material (SM).
Calling again $N(k,t)$ the number of vertices with degree $k$ at time $t$, its recurrence relation (see details in our SM) can be written as  $N(k,t+1) = N(k,t) \left(1 - \frac{k}{3t}\right) + N(k-2,t)\frac{k-2}{3t} + O\left(\frac{1}{t^2}\right)$ for $k>2m_{tri}$, and $N(2m_{tri}, t+1) =  N(2m_{tri}, t) \left(1 - \frac{2m_{tri}}{3t}\right) + 1 + O\left(\frac{1}{t^2}\right)$ for $k=2m_{tri}$.
In order to obtain an expression for $P(k)$, one supposes that  $\frac{N(k,t+1)}{t+1}=\frac{N(k,t)}{t}$ for large $t$. Therefore, one gets $P(2m_{tri}) = \frac{3}{3+2m_{tri}} $ and $P(k) = P(k-2)\frac{k/2+1}{k/2 + 1.5}$, which ultimately gives $
P(k)= \frac{3}{3+2m_{tri}} \prod_{l=m_{tri}+1}^{k/2} \frac{l-1}{l+1.5}$, or alternatively
\begin{equation}
    P(k) = \frac{3}{3+2m_{tri}} \frac{\Gamma(k/2)\Gamma(m_{tri} + 2.5)}{\Gamma(m_{tri})\Gamma(k/2+2.5)}\sim k^{-2.5},
    \label{eq:powlaw}
\end{equation}
where $\Gamma$ is here the gamma function.
The power law scaling predicted by Eq. (\ref{eq:powlaw}) fits remarkably well the numerical data of Figure 1 (see the dotted, dashed-dotted, and dashed black lines in the bottom-left panel).

As for $P(k_l)$, one calls again $N_e(k_l,t)$ the number of edges participating in $k_l$ triangles at time $t$. The recurrence relation for $N_e(k_l,t)$ (see details in our SM) is $N_e(k_l,t+1) =  N_e(k_l,t) \left(1 - \frac{k_l}{3t}\right) +
N_e(k_l-1,t)\frac{(k_l-1)}{3t} + O\left(\frac{1}{t^2}\right)$ for $k_l>1$, and $N_e(1,t+1)  =  N_e(1,t) \left(1 - \frac{m_{tri}}{3m_{tri}t}\right) + 2m_{tri} + O\left(\frac{1}{t^2}\right)$ for $k_l=1$. Imposing that $\frac{N_e(k_l,t+1)}{t+1}=\frac{N_e(k_l,t)}{t}$ for large $t$, one gets an equation for $P(k_l)$ which reads as $P(k_l) = P(k_l-1)\frac{k_l-1}{k_l+3}$, with $P(1) = \frac{3}{4} $. The solution is
\begin{equation}
    P(k_l)=\frac{3}{4} \prod_{l=2}^{k_l} \frac{l-1}{l+3} =\frac{3}{4}\frac{4!}{k_l(k_l+1)(k_l+2)(k_l+3)}\sim k_l^{-4}.
    \label{expolaw}
\end{equation}
The power-law function (\ref{expolaw}) is reported as  a dashed line in panel (b) of Fig. \ref{fig2}, and one can observe that the fit is, once again, extremely good.

\begin{figure}
\centering
\includegraphics[width=0.48 \textwidth]{{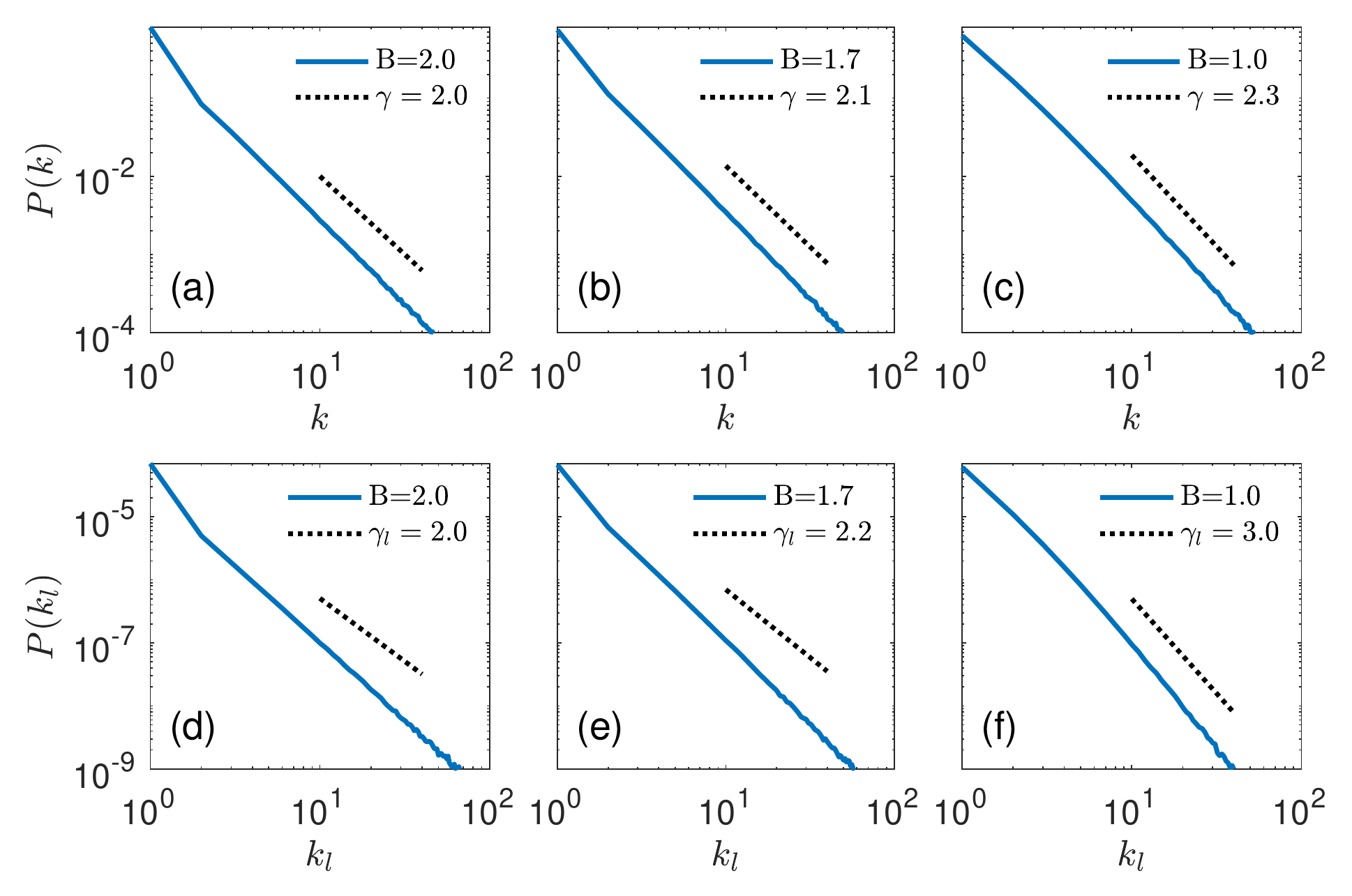}}
\caption{{\bf The mixed model.}  The distributions $P(k)$ [panels (a-c)] and $P(k_l)$ [panels (d-f)] obtained by growing
networks of size $N = 50,000$ with the mixed model, at three different values of $B$ (reported in each panel's top-right corner). The data (blue lines) refer to an average over 100 different realizations of the growth process.  Dotted lines report, for comparison, the scaling exponents predicted by Eq. (\ref{eq:exponents}). \label{mixmodel}}
\label{fig3}
 \end{figure}

Finally, our study can be extended and generalized to a mixed model, through which it is possible to effectively imprint {\it any desired} power-law scaling in the triangular structure of the network. To do so, we consider the case in which the probability that the node added at time $t$ selects the specific link $(i,j)$ to form its connections  is $P_{ij}(t)=A\frac{2}{N_l(t-1)} + B\frac{3k_{ij}(t-1)}{2\sum_{ij}k_{ij}(t-1)}$, for some constants $A$ and $B$. Here, $B$ is non-negative, $N_l(t)\sim2m_{tri}t$ is the number of links at moment $t$ and $\sum_{ij}k_{ij}(t-1)=3m_{tri}t$ is the sum of the generalized degrees of all edges. Notice that $A=0$ and $B=\frac{2}{3}$ ($A=\frac{1}{2}$ and $B=0$)  recovers the preferential (non preferential) case discussed above.

From the constrain that the sum of all probabilities must be equal to $1$, it follows that $A$ and $B$ must obey $2A + \frac{3}{2}B = 1$, so that
$A=1/2-3/4B$. Furthermore, $P_{ij}(t)$ must be non-negative for all $ij$ (also those for which  $k_{ij}=1$), and this gives the following bounds for $A$ and $B$:
$-1 \leq A \leq  \frac{1}{2} $ and $0\leq B \leq 2$.
Under these conditions, one can analytically demonstrate (see our SM for details) that, for strictly positive $B$ values the resulting networks display scale-free distributions $P(k)\sim k^{-\gamma}$ and $P(k_l)\sim k_l^{-\gamma_l}$ with exponents given by

\begin{equation}
    \gamma=1+\frac{1}{A+B}=1+\frac{4}{2+B}, \ \ \ \ \ \ \ \gamma_l=1+\frac{2}{B}.
    \label{eq:exponents}
\end{equation}

For $B=0$, one has instead (see again our SM for full details) $P(k_l)=\frac{2}{3^{k_l}}$.
Eq. (\ref{eq:exponents}) implies that $\gamma$ values are between $2$ ($B=2$) and $3$ ($B=0$), whereas $\gamma_l$ is equal to $2$ for $B=2$, and tends to infinity as $B$ tends to $0$. On its turn, this means that choosing $B$ between $1$ and $2$, $\gamma$ and $\gamma_l$ can be  pre-selected {\it ad libitum} between $2$ and $3$, i.e. the imprinted structures of links and triangles feature well defined mean values of the degrees, but unbounded fluctuations as the system grows in size.

In Figure \ref{fig3} we report $P(k)$ and $P(k_l)$ for three distinct values of $B$. It is seen that the fit between analytic predictions and numerically generated data is always very good. Moreover, the Figure demonstrates that our method constitutes a highly general scheme by means of which one can construct, in a fully flexible way, ensembles of synthetic complexes displaying any desired statistical properties [from the condition of panels (a,d) in Figure 3 featuring  a super scale-freeness - where even the mean degrees diverge in the thermodynamic limit-, to any milder condition which characterizes in fact many networks from the real world].

In summary, complex networks encode the basic architecture of social, biological and technological networks, touching upon the most crucial challenges of modern science – from the spread of epidemics in social networks \cite{vespi,baru1} to the resilience of our eco-systems and critical infrastructure \cite{baru2}.  In the context of pair-wise interactions, the most natural measure of centrality is a node’s individual degree, capturing its potential dynamic impact on the system \cite{baru3}. The discovery that most real world networks exhibit extreme levels of degree heterogeneity was disruptive – indicating that networks are highly centralized, with a potentially disproportionate role played by a small fraction of their components \cite{barabara,barabara2}.
As we deepen our investigation into the interaction patterns of complex systems it becomes increasingly clear, however, that higher order structures, beyond pair-wise interactions, underlie much of the observed richness of real-world networks. Hence, we seeked the fundamental rules that prescribe centrality, and govern its distribution, in a simplicial complex environment. A natural measure is the number of complexes, here triangles, that a  link participates in. Indeed, a simplicial complex represents a potentially functional unit, such as a collaboration of a social team \cite{patania}, or a trio of interacting biochemical agents \cite{Uzi,kuzmin}. A component that is part of many such complexes is, therefore, likely central in the functionality of the system.
Several growing network models have been introduced and studied, which help exposing the roots of simplicial complex heterogeneity, shedding light on the emergence of centrality beyond the degree distribution. As we seek to understand the behavior of complex systems, their resilience and dynamic functionality, we hope that our insight into the microscopic processes of their formation can provide meaningful macroscopic insights.

ISN acknowledges partial support from the Ministerio de Econom\'ia, Industria y Competitividad of Spain under project FIS2017-84151-P.

\begin{widetext}
\appendix
\section{SUPPLEMENTAL MATERIAL}
\subsection{The scripts of the codes}

In the following subsections we give the MatLab scripts for all generating algorithms used in our study.

\subsubsection{The non preferential attachment case}

\begin{verbatim}
function A=netsimplicial_random(N,ntri)
mlinks=2*ntri;
N0=mlinks+1;
A0=sparse(ones(N0,N0));
A0=A0-diag(diag(A0));
A(1:N0,1:N0)=sparse(A0);

for n=N0+1:N
  [i,j] = find(triu(A,1)>0);%List of links at
  %time n
    A(n,n)=0;
    %Step 1: Pick ntri random links from the
    % existing ones
    l=length(i); %Number of links
    isw=0;
    while(isw==0)
        m=randi(l,ntri,1);%ntri random positions
        %in the list of l links
        v=[i(m) j(m)];
        if(length(unique(v))==2*ntri)
            isw=1;
        end
    end
    for k=1:ntri
        A(n,i(m(k)))=1;
        A(i(m(k)),n)=1;
        A(n,j(m(k)))=1;
        A(j(m(k)),n)=1;
    end
end
\end{verbatim}

\subsubsection{The preferential attachment case}

\begin{verbatim}
function A=netsimplicial_preferential(N,ntri)
mlinks=2*ntri;
N0=mlinks+1;
A0=sparse(ones(N0,N0));
A0=A0-diag(diag(A0));
A(1:N0,1:N0)=sparse(A0);

Dij=A.*A^2;%each non diagonal element of this
%matrix returns the number of triangles the
%edge ij is forming

TDij=triu(Dij);
[I J]=find(TDij>0);
K=find(TDij>0);
plink=[I J full(TDij(K))];

pcum=[0 ;cumsum(plink(:,3)./sum(plink(:,3)))];
%accumulated probability vector of each link
%according to its participation in triangles

for n=N0+1:N
    A(n,n)=0;
    l=length(plink); %Actual number of edges

    isw=0;
    while(isw==0)
        dummy=rand(ntri,1);%Get ntri random
        %numbers
        diffs=pcuM^{e}-dummy;
        temp=diff(sign(diffs),1,2);
        [row,idx]=find(temp);% Find ntri
        %links whose accumulated
        %probability is larger than a value
        %picked at random
        if(length(unique(plink(idx,1:2)))==2*ntri)
            isw=1;
        end
    end
    for k=1:length(idx)
        inode=plink(idx(k),1);
        jnode=plink(idx(k),2);
        A(n,inode)=1;
        A(n,jnode)=1;
        A(inode,n)=1;
        A(jnode,n)=1;

        %Update the matrix Dij
        Dij(n,inode)=1;
        Dij(n,jnode)=1;
        Dij(inode,n)=1;
        Dij(jnode,n)=1;
        Dij(inode,jnode)=Dij(inode,jnode)+1;
        Dij(jnode,inode)=Dij(jnode,inode)+1;
        plink(l+1,:)=[n inode 1];
        plink(l+2,:)=[n jnode 1];
        plink(idx(k),3)=Dij(inode,jnode);
        l=length(plink);
    end

    %Update the probability of each link
    pcum=[0 ;cumsum(plink(:,3)./sum(plink(:,3)))];

end
\end{verbatim}

\subsubsection{The mixed case}

\begin{verbatim}
function [A,telaps]=netsimplicial_mixed(N,ntri,parA,parB)

N0=2*ntri+1;
A0=sparse(ones(N0,N0));
A0=A0-diag(diag(A0));
A(1:N0,1:N0)=sparse(A0);



Dij=A.*A^2;

TDij=triu(Dij);
[I J]=find(TDij>0);
K=find(TDij>0);
plink=[I J full(TDij(K))];
L=size(plink,1);
pcum=[0 ;cumsum(parA*ones(L,1)./sum(plink(:,3))+parB*plink(:,3)./sum(plink(:,3)))];

for n=N0+1:N
    A(n,n)=0;
    l=length(plink);

    isw=0;
    while(isw==0)
        dummy=rand(ntri,1);
        diffs=pcum'-dummy;
        temp=diff(sign(diffs),1,2);

        [row,idx]=find(temp);
      if(length(unique(plink(idx,1:2)))==2*ntri)
            isw=1;
        end
    end

    for k=1:length(idx)
        inode=plink(idx(k),1);
        jnode=plink(idx(k),2);
        A(n,inode)=1;
        A(n,jnode)=1;
        A(inode,n)=1;
        A(jnode,n)=1;

        Dij(n,inode)=1;
        Dij(n,jnode)=1;
        Dij(inode,n)=1;
        Dij(jnode,n)=1;
        Dij(inode,jnode)=Dij(inode,jnode)+1;
        Dij(jnode,inode)=Dij(jnode,inode)+1;
        plink(l+1,:)=[n inode 1];
        plink(l+2,:)=[n jnode 1];
        plink(idx(k),3)=Dij(inode,jnode);
        l=length(plink);
    end
    L=size(plink,1);
    pcum=[0 ;cumsum(parA*ones(L,1)./sum(plink(:,3))+parB*plink(:,3)./sum(plink(:,3)))];


end
\end{verbatim}
	
\subsection{Analytical results}

We here furnish more analytical results regarding our models.

The starting point is that, as the synthesized networks have transitivity coefficient $T=1$ (i.e. no links exist which do not form part of at least a triangle), the degree of each vertex is the number of triangles containing that vertex multiplied by 2, and one has that $N_v(k_v, t) = N(2k, t)$, where $N_v(k_v,t)$ ($N(2k, t)$) is the number of nodes participating in $k_v$ triangles (having degree $2k$) at time $t$. As a consequence one has that $P(k_v)=P(k)$, and is therefore entitled to concentrate on either one of such distributions, depending on which one finds the simpler analytical treatment.

\subsubsection{The non preferential attachment case}

Let $N_v(k,k_v,t)$ be the number of nodes with $k$ neighbours participating in $k_v$ triangles at time $t$. The rate equation is
\begin{eqnarray}
  \nonumber
  \frac{dN_v(k,k_v,t)}{dt}&=&-\frac{2m_{tri}}{\sum_kk N(k,t)}k N_v(k,k_v,t) \\ \nonumber &&+ \frac{2m_{tri}}{\sum_kk N(k,t)}(k-1) N_v(k-1,k_v-1,t) \\ &&  +\delta_{k,2m_{tri}}\delta_{k_v,m_{tri}}\\ \nonumber && \pm \dots ,
\end{eqnarray}
where the unwritten terms account for the formation of triangles from two or more linked edges. In the sequel, we assume that the chosen edges are not linked  (none of the nodes of a selected edge is linked to any of the nodes of another selected edge, which is always the case for $m_{tri}=1$). Hence, the number of edges and triangles related to a given node increases one by one. Only the new nodes entering the system have the number of edges ($2m_{tri}$) double than the number of triangles in which they are participating ($m_{tri}$). This way, for $k$ big enough, one has
\begin{equation}
  k N_v(k,k_v,t)\simeq k_v N_v(k,k_v,t).
\end{equation}
After summing over all values of $k$, one obtains an approximate equation for $N_v(k_v,t)$, the number of nodes participating in $k_v$ triangles:
\begin{eqnarray}
  \nonumber
  \frac{dN_v(k_v,t)}{dt}&\simeq &-\frac{2m_{tri}}{\sum_kk N(k,t)}k_v N_v(k_v,t) \\ \nonumber &&+ \frac{2m_{tri}}{\sum_kk N(k,t)}(k_v-1) N_v(k_v-1,t) \\ &&  +\delta_{k_v,m_{tri}}.
\end{eqnarray}
One then can proceed as in the case of the degree distribution $P(k)$ (see the main text), and seek a solution of the form $N_v(k_v,t)=tP(k_v)$:
\begin{equation}
  P(k_v)=\frac{k_v-1}{k_v+2}P(k_v-1)+\frac{2}{2+k_v}\delta_{k_v,m_{tri}}.
\end{equation}
The solution for $k_v\ge m_{tri}$ is
\begin{eqnarray}
  \nonumber
  P(k_v)&=&\frac{2m_{tri}(m_{tri}+1)}{k_v(k_v+1)(k_v+2)} \sim k_v^{-3}.
  \label{eq:ranPDeltav}
\end{eqnarray}
 Therefore, one has also that $P(k) \sim k^{-3}$.


\subsubsection{The preferential attachment case}

In order to obtain $P(k)$ and $P(k_l)$, one needs here to make use of the following Lemma from Ref. \cite{Chung2004}:

{\bf Lemma 1}
Suppose that a sequence $\{a_t\}$ satisfies a recurrence relation
$$
a_{t+1} = \left(1 - \frac{b_{t}}{t+t_1}\right)a_t + c_t,
$$
where $t_0$ and $t_1$ are arbitrary, positive, fixed, values.
Furthermore, suppose that $\lim_{t \to \infty} b_t = b > 0$ and $\lim_{t \to \infty} c_t = c > 0$.

Then $\lim_{t \to \infty} \frac{a_t}{t}$ exists and one has
$$
\lim_{t \to \infty} \frac{a_t}{t} = \frac{c}{1+b}.
$$

For instance (and indicating by $\mathbb{E} (\cdot)$ the expectation value), $P(k)$ for $k=2m_{tri}$ can be obtained by setting $b_t = b = \frac{2m_{tri}}{3}$ and $c_t=c=1$.
Application of Lemma 1 ensures that the limit $\lim_{t\to\infty}\frac{\mathbb{E}(N(2m_{tri},t))}{t}$ exists and is equal to $\frac{3}{3+2m_{tri}}$, leading to $P(1)=\frac{3}{3+2m_{tri}}$. On the other hand, $P(k)$ for $k>2m_{tri}$ is obtained assuming that $P(k-2)$ exists, and applying Lemma 1 again with $b_t=b=\frac{k}{3}$ and $c_t=\mathbb{E}(N(k-2,t))\frac{k-2}{3t}$  (i.e. taking $c = P(k-2)\frac{k-2}{3}$). Such a choice, indeed, entitles one to write a recurrence relation, and to obtain an explicit formula for $P(k)$:
\begin{equation}
    P(k) = \frac{3}{3+2m_{tri}} \frac{\Gamma(k/2)\Gamma(m_{tri} + 2.5)}{\Gamma(m_{tri})\Gamma(k/2+2.5)}\sim k^{-2.5},
\end{equation}
which coincides with Eq. (3) of the main text. Here, $\Gamma$ is the gamma function.

Furthermore, one can demonstrate that $P(k)$ is sharp, by the use of the following Theorem:

{\bf Theorem 1}
For any fixed $\varepsilon>0$ and $\delta > 0$ and for any large enough $t$, the difference between the number of vertices with degree $k$ at time $t$ and $P(k)t$ is smaller than $\varepsilon t$ with probability larger than $1 - \delta$.

The theorem can be proved by considering the martingale $X_l=\mathbb{E}(N(k,t)|\mathcal{F}_l)$
(where $\mathcal{F}_l$ is the $\sigma$-algebra generated by the probability space at time l). It is rather easy to show that $\mid X_{l+1}-X_{l} \mid$ is bounded by 4, and therefore the theorem follows from the Azuma-Hoeffding inequality \cite{hoeffding,azuma}.

When trying to obtain $P(k_l)$, one can encounter a problem in the case in which the triangles added at step $t$ have one or more common edges, so that the number of different edges added at time $t$ might not be equal to $2m_{tri}$. Denoting by $e(t, k_l)$ the number of added edges of degree $k_l$ at time $t$, the more accurate recurrence relation for $k_l>1$ is the following:
$$
\mathbb{E}(N_e(k_l,t+1)) =  \mathbb{E}(N_e(k_l,t)) \left(1 - \frac{k_l}{3t}\right) +
$$
$$
+\mathbb{E}(N_e(k_l-1,t))\frac{(k_l-1)}{3t} + O\left(\frac{1}{t^2}\right) + \mathbb{E}(e(t, k_l)),
$$
and for $k_l=1$ one has
$$
\mathbb{E}(N_e(1,t+1)) =  \mathbb{E}(N_e(1,t)) \left(1 - \frac{1}{3t}\right) + O\left(\frac{1}{t^2}\right) + \mathbb{E}(e(t, 1)).
$$

Using Theorem 1 (and some non trivial math, of which we omit the technical details), one can prove that $\lim_{t\to\infty} e(t,1)=2m_{tri}$ and $\lim_{t\to\infty} e(t,k_l)=0$ for $k_l>1$, so that one has $\lim_{t\to\infty} \mathbb{E}(e(t, 1))=2m_{tri}$ and $\lim_{t\to\infty} \mathbb{E}(e(t, k_l))=0$ for $k_l>1$.

Furthermore, one can prove another theorem:

{\bf Theorem 2}
The number of different edges divided by $t$ converges always to $2m_{tri}$.

It follows that $P(k_l)=\lim_{t\to\infty}\frac{\mathbb{E}(N_e(k_l,t))}{2m_{tri}t}$.

Finally, one can apply Lemma 1 with $b_t = b = \frac{1}{3}$ and $c=\lim 1+O\left(\frac{1}{t^2}\right) = 1$ to get that $P(1)=\lim_{t\to\infty}\frac{\mathbb{E}(N_e(1,t))}{2m_{tri}t}$ exists and is equal to $\frac{3}{4}$. For $k_l>1$, one assumes that $P(k_l-1)\lim_{t\to\infty}\frac{\mathbb{E}(N_e(k_l-1,t))}{2m_{tri}t}$ exists, and applies again Lemma 1 with $b_t = b = \frac{k_l}{3}$ and $c = \lim c_t=P(k_l-1)\frac{(k_l-1)}{3t}$. Hence, $P(k_l)=\lim_{t\to\infty}\frac{\mathbb{E}(N_e(k_l,t))}{2m_{tri}t}$ exists and is equal to $P(k_l-1)\frac{k_l-1}{k_l+3}$. From such a recurrence relation, one finally obtain the explicit formula for $P(k_l)$:

\begin{equation}
    P(k_l)=\frac{18}{k_l(k_l+1)(k_l+2)(k_l+3)} \sim k_l^{-4},
\end{equation}
which is identical to Eq. (4) of the main text.

Finally, since the degree of the vertex is the number of triangles containing this vertex multiplied by 2, one has $N_v(k_v, t) = N(2k, t)$ and as a consequence one obtains
\begin{equation}
    P(k_v) = \frac{3}{3+2m_{tri}} \frac{\Gamma(k_v)\Gamma(m_{tri} + 2.5)}{\Gamma(m_{tri})\Gamma(k_v+2.5)}\sim k_v^{-2.5}.
    \label{piropiropiro}
\end{equation}

\subsubsection{The mixed case}

The recurrence relation for $N_v(k_v,t)$ (the number of vertices participating in $k_v$ triangles at time $t$) is

$$
\mathbb{E}(N_v(k_v,t+1)) =  \mathbb{E}(N_v(k_v,t))\left(1 -A\frac{m_{tri}}{t}- (A+B)\frac{k_v}{t}\right) +
$$
$$
 + \mathbb{E}N_v(k_v-1,t)\left(A\frac{m_{tri}}{t}+(A+B)\frac{(k_v-1)}{t}\right) + O\left(\frac{1}{t^2}\right).
$$

For $t>0$ and $k_v=m_{tri}$ one has
$$
\mathbb{E}(N^{v}_{t+1}(m_{tri}))=\mathbb{E}(N_v(m_{tri},t)) \left(1 - A\frac{m_{tri}}{t} - (A+B)\frac{m_{tri}}{t}\right) + 1.
$$

Furthermore, the use of Lemma 1 gives

$$
P(m_{tri}) = \frac{1}{1+2Am_{tri}+Bm_{tri}}
$$
and
$$
P(k_v)=P(k_v-1)\frac{Am_{tri}+(A+B)(k_v-1)}{1+Am_{tri}+(A+B)k_v}.
$$
Therefore, one has
\begin{equation}
P(k_v) =\frac{1}{1+2Am_{tri}+Bm_{tri}}\prod_{l=m_{tri}+1}^{k_v} \frac{Am_{tri}+(A+B)(l-1)}{1+Am_{tri}+(A+B)l}.
\end{equation}
or
$$
P(k_v) =\frac{1}{1+2Am_{tri}+Bm_{tri}}
\cdot\prod_{l=m_{tri}+1}^{k_v} \frac{\frac{Am_{tri}}{A+B}+l-1}{\frac{1+Am_{tri}}{A+B}+l} =
$$
$$
=\frac{1}{1+2Am_{tri}+Bm_{tri}}\cdot\frac{\Gamma(k_v+\frac{Am_{tri}}{A+B})\Gamma(m_{tri} + \frac{1+Am_{tri}}{A+B}+1)}{\Gamma(m_{tri}+\frac{Am_{tri}}{A+B})\Gamma(k_v+\frac{1+Am_{tri}}{A+B})+1)} \sim k_v^{-\gamma_v},
$$
where $\gamma_v=1+\frac{1}{A+B}$.

On the other hand, one can consider $N_l(k_l,t)$, the number of links with degree $k_l$ at moment $t$. Its recurrence equation is
$$
\mathbb{E}(N_l(k_l,t+1)) =  \mathbb{E}
 N_l(k_l,t) \left(1 - A\frac{1}{t} - B\frac{k_l}{2t}\right) +\mathbb{E}N_l(k_l-1,t)\left(A\frac{1}{t} + B\frac{k_l-1}{2t}\right) + O\left(\frac{1}{t^2}\right).
$$

For $k_l=1$ one has
$$
\mathbb{E}(N_l(1,t+1)) =\mathbb{E}(N_l(1,t)) \left(1 - A\frac{1}{t} - B\frac{1}{2t}\right) + 2m_{tri} + O\left(\frac{1}{t^2}\right).
$$

One can apply again Lemma 1, and obtain
$$
P(1) = \frac{2m_{tri}}{1+A+\frac{1}{2}B} , \ \ \ \
P(k_l)=P(k_l-1)\frac{2A + B(k_l-1)}{2+2A + Bk_l}.
$$
Therefore, one gets
\begin{equation} \label{eq:edge_distr_prod}
P(k_l) = \frac{2m_{tri}}{1+A+\frac{1}{2}B} \prod_{l=2}^{k_l} \frac{2A + B(l-1)}{2+2A + Bl}.
\end{equation}

Let us consider separately the cases $B=0$ (the fully non preferential case) and $B>0$.

For $B=0$, one has
$$
P(k_l) = \frac{2m_{tri}}{1+A} \prod_{l=2}^{k_l} \frac{2A }{2+2A}.
$$
Since $A=\frac{1}{2}$, one obtains
$$
P(k_l) = \frac{4m_{tri}}{3} \prod_{l=2}^{k_l} \frac{1}{3}=\frac{4m_{tri}}{3}\frac{1}{3^{k_l-1}},
$$
which fully coincide with the exponential scaling derived in the main text for the non preferential case.

On the other hand, when instead $B>0$, one has
$$
P(k_l) = \frac{2m_{tri}}{1+A+\frac{1}{2}B} \prod_{l=2}^{k_l} \frac{\frac{2A}{B} + l-1}{\frac{2}{B}+\frac{2A}{B} + l}=\frac{2m_{tri}}{1+A+\frac{1}{2}B}\frac{\Gamma(\frac{2+2A}{B}+2)\Gamma(\frac{2A}{B}+k_l)}{\Gamma(\frac{2A}{B}+1)\Gamma(\frac{2+2A}{B}+k_l+1)}\sim k_l^{-\gamma_l},
$$

where $\gamma_l=1+\frac{2}{B}$.
\end{widetext}
\bibliography{references}

\end{document}